
\input psfig
\vsize=9 truein
\hsize=6.5 truein
\parskip=0.2truecm
\parindent=.5truecm
\raggedbottom


  \def\refs{\hangindent=5ex\hangafter=1}
  \newcount\fcount \fcount=0
  \def\ref#1{\global\advance\fcount by 1 \global\xdef#1{\relax\the\fcount}}

\def\pp{\parshape 2 0truecm 15truecm .5truecm 14.5truecm}
\def\book #1;#2;#3{\par\pp #1, {\it #2}, #3}

%
\def\apjref#1;#2;#3;#4 {\par\pp#1, {\it #2}, {\bf #3}, #4. \par}
%
%

\def\rep #1;#2;#3{\par\pp #1, #2, #3}

\def\simlt{\lower.5ex\hbox{$\; \buildrel < \over \sim \;$}}
\def\simgt{\lower.5ex\hbox{$\; \buildrel > \over \sim \;$}}


\centerline { {\bf BARYONIC DARK MATTER}\footnote{$^\dagger$}{To be published
in the proceedings of the 1993 Les Houches Summer School on Theoretical Physics
on {\it Cosmology and Large--Scale Structure}}}
\bigskip
\centerline {Joseph Silk}
\medskip
\centerline {Departments of Astronomy and Physics, and Center for Particle
Astrophysics}
 \centerline {University of California, Berkeley, CA 94720}

\medskip

{\it ``La theorie c'est bien, mais ca n'empeche pas d'exister.''}

\centerline {\it Emile Charcot}
\bigskip
{\bf Abstract}

In the first two of these lectures, I present the evidence for baryonic dark
matter and describe
possible forms that it may take.  The final lecture discusses formation of
baryonic dark matter, and sets the cosmological context.

1.  {\bf INTRODUCTION}

The nature of the dark matter represents one of the major unsolved problems
in astrophysics.  In fact, there are really two dark matter problems:  the dark
matter in the halo and the dark matter that is the predominant contributor
to $\Omega.$  Baryonic dark matter is a plausible candidate for halo dark
matter,
but whether it is responsible for $\Omega$ is controversial.  If $\Omega=1,$ it
is most unlikely that baryons predominate, but if $\Omega \sim 0.1,$ the
situation is less clear.

The uncertainty in the masses of baryonic and non--baryonic dark matter
candidates is huge, and should give any experimentalist serious grounds for
hesitation before embarking on a finely tuned search for dark matter.  This
uncertainty has never deterred theorists:  on the contrary, it has inspired
them to come up with a plethora of candidates.  A convenient categorization of
dark matter candidates divides the contenders into two regimes, that of
particle physics and that of astrophysics.  The particle physics--motivated
non-baryonic candidates are weakly interacting particles, generally but not
invariably massive that are generically classified as WIMPs, for $w$eakly
$i$nteracting $m$assive $p$articles.  The astrophysically motivated baryonic
candidates are generally massive and are referred to as MACHOS, for $m$assive
$a$strophysical $c$ompact $h$alo $o$bjects.  MACHOs may of course exist beyond
galactic halos, but the evidence is less persuasive.

Generally, one can compute the cosmological abundance of WIMPs but there is no
proof of their existence.  The most compelling WIMP is the lightest
supersymmetrical particle, and provides the best candidate for the dark matter
that contributes to $\Omega=1,$ if indeed $\Omega$ is unity (or even
larger).  By contrast, MACHO candidates are known to exist, and these lectures
will describe
the evidence for MACHOs as the dark halo matter.  Arguments will be reviewed
that both support MACHOs as halo dark matter, and go the additional step of
asserting that all dark matter may be baryonic (Lecture 1).

These lectures were particularly timely.  Shortly after they were given, and
before being written up, evidence for MACHO candidate detections was
reported in two independent experiments.  While these results support the
idea that halo dark matter is baryonic, they leave wide open, at present, any
inferences about the nature of the MACHOs.  Thus in Lecture 2, I describe the
possible forms of baryonic halo dark matter that span the range of compact
stellar remnants, brown dwarfs and even cold gas clouds.  The third lecture
sets the cosmological context, and describes how a viable cosmological model
may be constructed that consists exclusively of baryonic dark matter.

2.  {\bf THE EVIDENCE FOR BARYONIC DARK MATTER}

{\it 2.1.  Primordial Nucleosynthesis}

The strongest argument that there is a substantial amount of baryonic dark
matter, exceeding luminous baryonic matter, comes from primordial
nucleosynthesis.  The concordance of the predictions of $^4He,$ $^3He,$ $^2H$
and $^7Li$ abundances, together with the removal of uncertainty in the numbers
of neutrino species by the $Z^0$ decay width and improved measurement
of
neutron half--life, bounds $\Omega_B.$  Indeed, the major uncertainty arises
from the adopted Hubble constant.  The upper limit on $\Omega_B$ is primarily
from $^4He$ and $^7Li,$ and the lower bound from $^2H$ and $^3He.$  The
combined result is \ref\waletal $[\waletal]$ $\Omega_Bh^2 = 0.0125 (\pm
0.0025);$ $95\% $ confidence limit.  Since
luminous baryons contribute $\Omega^{lum}_B \approx 0.007,$ I conclude that
between $30\%$ and $90\%$ of baryons are dark.

These bounds contain hidden assumptions about galactic chemical evolution.  If
halo dark matter consists of stellar remnants, one may have to reevaluate these
limits.  Indeed, there are already indications from population II abundance
studies that lithium may have undergone convective burning at an early stage
of galactic chemical evolution \ref\pin$[\pin].$  For example, the ratio
${^6Li/ ^7 Li}
\approx {1\over 20}$ in HD 84937, an old halo star with $[{Fe/ H}] = -2.4$
\ref\smil$[\smil],$ implies that old
halo stars have certainly destroyed $^6Li$ by convective  burning
\ref\steetal$[\steetal].$  Cosmic ray
spallation of $(\alpha,\alpha)(p, ^7Li)$ and $(p,CNO)$
$(p, ^7Li,...)$ undoubtedly produces some
$^7Li,$  as evidenced by the quadratic correlation of $[Be/H], [B/H]$ with
$[{Fe/ H}]$ \ref\walb$[\walb].$  This further complicates use of $^7Li$ as a
tracer of
$\Omega_B.$
 If $^6Li$ was destroyed, it becomes more plausible that convection to a
slightly greater depth could have destroyed a substantial amount of $^7Li.$

The discovery of MACHO candidates (section $2.5$) means that the dark halo may
consist of
stellar relics.
The burning of $^2H$ would generate
$^3He$ that could also be locked up, allowing a large astration, or stellar
destruction, factor of
$^2H$ and therefore a lower value of $\Omega_B.$  If the MACHOs formed from
intermediate mass stars, there wouls also be some helium production that would
also lower the inferred primordial helium abundance, and therefore also
$\Omega_B.$
The MACHO discovery
argues against a large astration factor, since it indicates a value of
$\Omega_B$ that is at least as large as that allowed by the conventional bound.
 Indeed halo MACHOs, if stellar remnants, favour a value of $\Omega_B$ that
could substantially exceed the canonical bound, by virtue of the lock-up of
what may be a high primordial helium abundance.

{\it 2.2.  Evidence from Rotation Curves}

Rotation curves of spiral galaxies show that halos constitutes some 90 percent
of the mass of a galaxy and are dark.  The local mass--to--luminosity ratio is
$\simgt 2000$ in the $V$-band, and $\simgt 64$ at $K,$ for edge--on spirals
\ref\skr$[\skr].$  This
precludes ordinary stars from being more than a few percent of the dark halo
mass.  Even a finely tuned mass function rising below $0.1 M_{\odot}$ would
produce too much light, if the stars were hydrogen burning $(M>0.08 M_\odot).$
Specific alternative baryonic candidates are discussed below.  However, a
simple argument derived from HI observations of at least some rotation
curves suggests that baryonic dark matter is a serious contender for explaining
halo dark matter.

Consider the curious case of the dwarf galaxy  DDO 154.  This dwarf spiral has
a flat rotation curve that extends to 15 disk scale lengths.  The measured HI
column density has the same radial profile $(N \propto r^{-1})$ as the
inferred projected dark matter surface density over this range
\ref\carf$[\carf].$  The HI
contributes about 10 percent of the total surface density, and accounts for
about 30 percent of the measured rotational velocity.  The logical contender
for the dark matter is a component that is associated with the halo HI.  This
points towards baryonic dark matter and possibly even cold gas clouds, as a
dominant form of dark matter.  To what extent cold gas can exist in
substantial amounts in
the halo will be discussed below.

Another characteristic of DDO 154 and other dwarf spirals is that the rotation
curve observations provide unambiguous evidence that the extended dark matter
halos have cores of finite extent, typically several kiloparsecs. However
halo formation by hierarchical clustering of collisionless dark matter
particles
in high resolution numerical simulations demonstrates that the density
increases to below the resolution limit \ref\dubc\ref\warq $[\dubc,\warq].$
Any core radius is less than the resolution scale  of $\sim\rm 1 \, kpc.$
While one can imagine extreme mass loss via supernova--driven winds imprinting
a scale on the dark matter halo, a more plausible alternative is simply a
dissipative BDM halo \ref\moore$[\moore].$ This is the simplest way to generate
a halo with a finite core.

{\it 2.3.  Dark Halos May Be Flattened}

One signature of baryonic dissipation is an oblately flattened halo.
Collisionless collapse in hierarchical clustering tends to form prolate halos.
There are indications that dark halos may be flattened.  Studies of polar ring
galaxies probe the halo potential along the minor axis and provide evidence of
flattening comparable to an E6 galaxy (or 4:10) \ref\sacs$[\sacs].$  Since
polar ring galaxies
form from mergers, one might expect that isolated galaxies could be more
flattened.

 HI disk warps provide indirect evidence of flattening.  The
argument here is one of persistence of the non-axisymmetric instability.  An
oblate spheroidal halo damps out the warp modes over a time-scale
$\sim \epsilon^{-1}t_p$ where $\epsilon$ is the flattening (ratio of semiminor
to semimajor axes) and $t_p$ is the precession
time-scale \ref\nelt $[\nelt].$  Survival over the age of the galaxy requires
$\epsilon \simlt 0.3.$

Lopsided HI disks with circular velocity fields \ref\begb\ref\sana\ref\rup
$[\begb,\sana,\rup]$ also provide evidence for
dark baryonic matter.  These are not easily understood even if the gas has
recently arrived:  the simplest explanation for lopsidedness in outer HI disks
is
that the disk is dark-matter dominated.  This matter must be baryonic in order
to be in the disk.

 Stellar kinematics also favor a flattened population II \ref\whi\ref\binm
$[\whi,\binm].$
Presumably a gaseous halo that formed contemporaneously with population II
would be even more flattened.

Simulations of hierarchical clustering in a cold dark matter--dominated
universe preferentially form prolate halos $[\dubc].$ This result is not in
agreement with evidence on
the distribution of shapes for elliptical galaxies.  This may suggest that the
simulations are not providing the correct picture since one would expect
stars to have formed early in the collapse. A prolonged episode of star
formation
would result in dissipation and disk formation.  The stars should therefore
dynamically track the dark matter.

{\it 2.4.  Continuity With Population II}

There are two observationally motivated arguments for believing that dark halos
are baryonic.  The disk-halo ``conspiracy'' is the apparent dependence of both
the amplitude and shape of the galaxy rotation curve at large galactocentric
radii (that is, in the halo-dominated regime)  on the luminous content of the
galaxy (that is, the disk).  Flat rotation curves are formed for normal, spiral
galaxies, with disk scale lengths of 3-4 kpc and maximum rotational velocities
of $100-200\rm \, km\, s^{-1}.$  However, luminous galaxies, with large
rotational velocities, generally have declining rotation curves, and dwarf
galaxies have
rising rotation curves \ref\casv$[\casv].$  In the galaxies with flat rotation
curves, the disk,
which dominates at small radii, and the halo, which dominates at large radii,
contribute an approximately equal rotational velocity.  Evidently the outer
rotation curve, which samples the galaxy halo, is closely coupled to the inner,
baryon-dominated galaxy.  This situation might naturally arise if the halo is
baryonic and formed shortly before the disk formed.

Stellar populations in the halo also show a continuity within population II.
The stars at large radii and of extremely low metallicity are drawn from a
similar narrow range of stellar masses and show similar abundance patterns to
population II stars in the inner galaxy.  The best nearby laboratories for
studying the oldest stars in the halo are globular star clusters. These systems
 are
inferred to contain a substantial fraction of mass in the form of white dwarfs
and neutron stars.  Perhaps 30 percent of a globular cluster may be in white
dwarfs, as determined by dynamical modelling, and at  least 1 percent in
neutron  stars is required to account for the millisecond pulsar population of
spun--up neutron stars in binaries.  The mass--to--luminosity ratio of an
elliptical galaxy $(\sim 10h)$ is several times larger than that of a globular
star cluster, and this may be in part due to a larger fraction of white dwarfs.
If the initial stellar mass function is sufficiently steep, globular clusters
which
have short central relaxation times would be depleted in low mass stars
relative to an elliptical.  However there is little indication that globular
clusters with the longest core relaxation times have steeper initial mass
functions.  While this effect  could conceivably account for the higher
mass--to--luminosity ratio in ellipticals, if these are low mass
star--dominated,
the mass--to--light ratio falls far short of that required in dark halos.  It
is an intriguing coincidence  that saturating the
lower limit on halo stellar dark matter, where
locally ${M\over L_v} \simgt 2000,$  could provide a critical density in the
same material, if it were uniformly distributed relative to the luminous
density of galaxies.

Baryonic dark matter may amount to $\Omega_B \sim 0.1 (h=0.4)$ or be as low as
$\Omega_B = 0.01 (h=1).$  It certainly exceeds the stellar contribution,
$\Omega_\ast \approx 0.007.$  Galaxy halos coincidentally span the range
where this dark matter
could be entirely baryonic.  The continuity argument suggests that halos are
the natural site for the baryonic dark matter.  Even galaxy clusters, where
gas and stars may dominate the mass, contribute no more than $\Omega_B \approx
0.1.$  Globally, only 90 percent of the mass of a spiral galaxy, halo included,
is dark. By dark, I mean that this material is not in any identifiable
form. Of course, there are white dwarfs and other stellar remnants that
cumulatively are dark and add up to a dominant fraction, 80 percent or more, of
amn old stellar population. There is dark matter over and above this near the
luminous peripheries of spiral galaxies.

The unknown dark fraction is not a huge extrapolation from the $\sim 20 $
percent in globular clusters
and $\sim 50$ percent in ellipticals that such population synthesis modelling
requires.
Thus somewhat more extreme pregalactic star formation could have produced the
requisite dark matter fraction.  Precisely what form the stellar dark matter
may take is the subject of section $3.$

{\it 2.5.  Gravitational Microlensing  Experiments}

Results from two experiments that find strong evidence for the existence of
MACHOs were reported in October, 1993 \ref\alcet\ref\aubet $[\alcet,\aubet].$
The technique used is gravitational
microlensing.  If a MACHO passes very close to the line--of--sight from Earth
to a distant star, the gravity of the otherwise invisible MACHO causes bending
of the starlight and acts as a lens.  For halo MACHOs, the star splits into
multiple images that are separated by a milliarcsecond, far too small to
observe.  However, the background star temporarily brightens as the MACHO moves
across the line--of--sight in the course of its orbit around the Milky Way
halo. The brightening may be by a magnitude or more, which should easily be
detectable.

The microlensing event has some unique signatures that distinguish it
from a variable star.  It should be symmetrical in time, achromatic, and should
occur only once for a given star.  There are two major difficulties with this
experiment.  First, the
microlensing events are very rare. Only about one background star in two
million will be microlensed at a given time.  Secondly, many stars are
intrinsically variable. Studying such rare events may uncover new types of
hitherto unknown variable stars.

To overcome the low probability of a microlensing event towards the LMC, about
$5\times 10^{-7},$ the
experiments were designed to monitor some ten million stars in the Large
Magellanic
Cloud.  One group, the EROS collaboration of French astrophysicists, utilized a
total
of more than 300 ESO Schmidt plates taken of the LMC over a  3  year period
with red or blue filters. The second group is a US--Australian collaboration
that utilizes the 50 inch
telescope at Mt. Stromlo, dedicated to the MACHO search, in
conjunction with the largest CCD camera in the world built for astronomical
use.

An analysis of about seven million stars revealed a total of four  events that
displayed the
characteristic microlensing signatures, with  event durations of between 20 and
40 days.
The duration of the microlensing event directly measures the mass of the MACHO,
with some uncertainty because of the unknown transverse velocity of the MACHO
across the line--of--sight.  The duration of the event is simply the time for
the MACHO to cross the Einstein ring radius.  The  Einstein ring radius is
approximately equal to the geometric mean of the Schwarzschild radius of the
MACHO and the distance to the MACHO. The MACHO is typically
at a halo core radius, which is a sizeable fraction  of the distance (55 kpc)
to the LMC.

  Much more data remains to be analyzed by the two groups.
The MACHO interpretation, if correct, should result in more events that are
distributed according to the expected distribution both of amplifications and
of the properties of the background stars. In the meantime, one can speculate
about the implications.  The four events
correspond to MACHO masses of 0.1 to 0.4 M$_\odot$ with a factor of 3 or so
uncertainty.  The EROS experimenters are also performing a CCD search that is
sensitive to timescales between 30m and 24 h, and therefore to mass scales
between $10^{-7}$ to $10^{-3}$ M$_\odot.$  As yet, no events have been found in
this mass range.

A third experiment, the OGLE  collaboration of Polish and U.S. astronomers
\ref\pacet $[\pacet],$ has studied 0.7 million
stars in the galactic bulge, where there is a higher microlensing probability
of detecting disk stars than halo MACHOs towards the LMC.  They reported
detection of a microlensing event corresponding to a mass of about
$0.3 \rm \, M_{\odot}.$  This approach will eventually provide confirmation of
the microlensing technique,
since one can predict a minimal expected rate of events from the known disk
stellar population.  By contrast, the halo would generate no MACHO events if
it does not consist of baryonic dark matter.

The rate detected appears to be low, perhaps by a factor of three,  relative to
what the MACHO model of dark halo matter
predicts. These conclusions are extremely tentative, and are sensitive to the
uncertain experimental efficiency and adopted halo model.
The possibility that the detections refer to a thick disk cannot be ruled out
if the thick disk contributes about as much as the stellar component of the
thin disk. However conventional thick disk observations require at most a six
percent contribution in terms of stellar surface density near the sun. These
experiments  certainly present the strongest evidence to date of dark matter
detection.   Unless there are perverse types of rare variable stars, MACHOS are
likely to
constitute a significant fraction of the dark halo.

  {\bf 3. THE POSSIBLE FORMS OF BARYONIC DARK MATTER}

{\it 3.1 A Star Formation Primer}

Star formation is a phenomenological theory.  We would like to be able to apply
this theory to the formation of BDM.  In this section, we review the current
status of the theory of star formation.  There are three critical ingredients
that are essential for understanding how stars form.  These are the initial
mass function of newly formed stars (IMF), the rate of star formation (SFR),
and the star formation effiency (SFE).

{\it 3.1.1  Initial Mass Function}

Most of our knowledge of the initial mass function comes from studies of stars
in the vicinity of the sun.  The IMF is defined as the total number of stars
per unit mass ever formed.  It is measured per square parsec perpendicular to
the galactic plane by counting field and open cluster stars of known distance.
Historically, the IMF is approximated by a Salpeter law over $0.1\rm\,
M_{\odot}$ to
$80 \rm\, M_{\odot},$
$${dn \over dm} \propto M^{-1 -x};\ \  x = 1.35.$$
Modern data shows that the IMF peaks at $0.3  \rm M_{\odot},$ and
declines towards lower masses.  At higher masses, the IMF gradually steepens.
It has been more accurately approximated as a log normal distribution by Miller
and Scalo \ref\mils $[\mils].$

The mass range over which one can observe the IMF outside the solar vicinity
is limited.  In regions of star formation, near infrared imaging has shown that
low
mass stars are present in numbers consistent with the Miller-Scalo IMF.  This
of course is essential for the question of how much mass is locked up in stars,
and in particular in low luminosity stars.  Studies of globular clusters do
not find evidence for a turn--over, and the numbers of stars continue to rise
to the observational limit of about $0.15  \rm M_{\odot}.$  In nearby galaxies,
where star formation occurs relatively quiescently, there are also no
indications of
deviations from the Miller--Scalo IMF at masses down to about $\sim 1
M_{\odot}.$

In the Milky Way, observations of the number of Wolf--Rayet stars suggest that
the IMF steepens as a function of increasing galactic radius
\ref\conv$[\conv].$  The budget of
ionizing photons, as measured by observations of radio HII regions, has been
used
to favour a top--heavy IMF in the inner spiral arms \ref\mez$[\mez].$

However, in starbursts, regions of
intense star formation activity,  both modelling and observational
indicators suggest that the IMF may vary with time and/or location. Arguments
for a top--heavy IMF, weighted towards massive stars, have been summarized by
Scalo \ref\scalo$[\scalo].$ The high
luminosity per unit of gas mass available to form stars suggests that the IMF
 is
truncated at the low mass end.  The best observational evidence  is for
M82, where spectroscopy  of the CO bands near $2 \mu$ \ref\rie $[\rie],$ the
high supernova rate and luminosity per unit gas mass \ref\doam $[\doam],$ and
the enhanced ratio of $K$--band luminosity to mass relative to that in the
nucleus \ref\les $[\les]$
all favor a supergiant--dominated starburst in the disk, driven by a top--heavy
IMF.

A further argument in favour of a top--heavy IMF during galaxy formation
comes  from the excessive enrichment  that is generated, relative to the
nucleosynthetic yield in the solar neighbourhood.
In galaxy clusters, the high abundance of iron in the intracluster medium,
about one--third of the solar iron abundance, has been interpreted as possibly
requiring a top--heavy IMF during the formation phase of the observed
ellipticals \ref\arnr\ref\renc $[\arnr,\renc].$ A Miller--Scalo IMF fails by a
factor of 10 in providing enough iron.

There is also some evidence of a deficiency in {\it very} massive stars in at
least some starbursts.
 The absence of massive stars
is
inferred from the paucity of ionizing photons as measured by the strength of
hydrogen recombination lines.
The primary theoretical argument for explaining this is due to
Wolfire and Cassinelli \ref\wolf $[\wolf],$ who note that in metal--rich
galactic nuclei, enhanced
radiation pressure due to small dust grains is more likely to limit accretion
onto
massive protostars than in lower metal abundance regions. This would be a
primary factor in limiting the upper limit on the IMF to a mass as low as
$\sim 30 \rm  M_{\odot}.$

There is no compelling theory that predicts stellar masses, let alone the IMF.
The characteristic stellar mass can be derived by simple dimensional arguments
that balance pressure and gravity to have a baryon number of
$({{\hbar c} \over {Gm_p^2}})^{3 \over 2}$ , equivalent to $10^{57}$ protons,
or to a solar mass.  This number is uncertain by at least two orders of
magnitude.
 Indeed, essentially the same argument has been used to derive the brown dwarf
mass
$(0.08 \rm  M_{\odot}),$ the Chandrasekhar mass $(1.4 \rm  M_{\odot}),$ and the
maximum
mass of a stable star $(\sim 100  \rm M_{\odot}).$

A cold, collapsing cloud will realistically form a transient sheet or filament
rather than collapse to a point \ref\lara$[\lara].$  It radiates freely during
the initial
collapse, and is unstable to fluctuation growth according to the Jeans
criterion.  The minimum fragment mass in  a cloud at temperature $T$ and
surface density $\mu$ is
$$M_{Jeans} \approx {{c_s^4} \over {G^2\mu}} = {{1.6 \left(T / 10
\rm K\right)^2 } \over{({\mu / 150} M_{\odot}\rm  pc^{-2})}}\rm  M_{\odot}.$$
A typical value for the surface density of molecular clouds on scales from
$0.1$ to $30\rm pc$ is $\mu = 150 \rm M_{\odot} pc^{-2}.$  The temperature is
in the range
$10\rm K - 50 K.$  The resulting Jeans mass spans the range observed for
molecular
cloud cores.

 Exactly how stars form depends on the continuing evolution and
subfragmentation of these cores.  Considerable amounts of magnetic flux and
specific angular momentum must be lost by the cores on the way to forming
stars.  Within the more massive cores, large numbers of stars form.  Numerical
hydrodynamical simulations cannot cope with the dynamical range in
density required to study star formation.  One has to resort to semi-analytical
arguments.

The resolution must depend on the initial conditions in the parent cloud.  How
does the cloud divide itself into stellar mass fragments?  The physics of cloud
collapse and evolution is complex.  It involves fragmentation, coalescence of
fragments, accretion by fragments, and binary captures.  It is not surprising
that the IMF may depend on environment, being different in the nuclei of
galaxies,
for example, from the IMF in lower density regions.  There are indications of
gradients of $\alpha$--nuclei abundances to iron in elliptical galaxies
\ref\worf $[\worf],$  and in the disk
of our galaxy \ref\gus $[\gus].$ These  can be interpreted in terms of IMF
variations.  However, this
invariably is not a unique explanation, as both supernova--driven mass loss
from galaxies and
accretion of primordial gas into galaxies can modify the abundance gradients.

The top--heavy IMF may arise as follows \ref\silkm$[\silkm].$  Interstellar
clouds grow by
coalescence and then orbit the galaxy.  Initially, magnetic support was
adequate to provide support against gravitational collapse, with ambipolar
diffusion of the field allowing some modest degree
of star
formation to proceed even in the low mass cores.  Because of the limited gas
reservoir, one might imagine that predominantly low mass stars  are formed in
these cores.  Spiral density waves provide a
non--circular component to the motion that progressively stimulates the
aggregation process.
After about an orbital time, $10^8 \rm yr$ or so, many clouds have grown to the
point
at which they are both Jeans unstable, and magnetically Jeans unstable.  The
massive  clouds now collapse on a free--fall time and
form stars of all masses.  Evidently, one has two modes of star formation.
During the prolonged, quiescent star--forming mode, low mass star formation
predominates.
Once cloud collapse begins in earnest, both massive and low mass stars form, in
the vigorous star--forming mode.

Now consider what may happen in the merger of a pair of gas--rich galaxies.
The greatly enhanced non--circular cloud motions should drive cloud growth by
aggregation on an unprecedented scale \ref\barh $[\barh].$  In this situation,
the vigorous
star--forming model dominates, as the clouds are rapidly driven to the edge of
collapse.  Hence a top--heavy IMF may arise naturally in regions of intense
turbulent motions of clouds as expected in a galaxy merger, and possibly also
during the process of galaxy spheroid formation.

{\it 3.1.2  Star Formation Efficiency}

Stars form in dense molecular cores that permeate the giant molecular cloud
complexes (GMC) \ref\lad$[\lad].$  The galactic molecular hydrogen, amounting
to about $2 \times 10^9  \rm M_{\odot}$ and comparable in mass to the atomic
hydrogen, is distributed in $\sim 1000$ of these cloud complexes.  The overall
SFE within the GMCs is a few percent \ref\myeetal$[\myeetal],$ but within
the most massive cores the SFE is 30
percent or more \ref\ladl$[\ladl].$, The cores represents a few percent of the
mass in the
cloud complexes.  The overall SFE is about 1 percent in the Milky Way disk.

One can understand the low SFE in terms of energy feedback once stars form.
Low mass as well as massive protostars are observed to have vigorous bipolar
outflows.  In addition,
low mass protostars are strong x--ray emitters.  The enhanced ionization
recouples the magnetic field that is undergoing ambipolar diffusion as slow
contraction  of the cloud cores occurs and low mass stars form. The increased
friction between ions and the molecular gas thereby provides an additional
magnetic pressure
source that resists collapse.  The bipolar outflows that are invariably
associated with the formation of stars inject a significant amount of momentum
and energy into the cold molecular gas, the bulk of which has not yet
condensed.

A crude measure of efficiency for supra--Jeans mass clouds is obtained as
follows \ref\silka$[\silka].$  Let typical
bipolar outflows be at velocity  $v_{out} \sim 100 - 200 \rm \, km\, s^{-1}$ in
molecular clouds
of linewidth $\Delta v \sim 1-3 \rm\, km\, s^{-1},$ where characteristic values
are
used.  If momentum is approximately conserved, one would expect that the
SFE is $\sim {{\Delta  v} / {v_{out}}},$ or of order one or two percent, over
the
time--scale over which the flows persist.  This might apply over a cloud
collapse time--scale, since this is of the same order ($\sim 10^6-10^7 \rm \,
yr$ at a density
$n=10-1000\rm \, cm^{-3}$) as the duration times estimated for many bipolar
outflows.  Hence feedback from  protostellar flows could account for the SFE
during the vigorous star--forming phase of
a molecular cloud, when gravitational collapse is underway.

  {\it 3.1.3 Star Formation Rate}

The star formation rate in galaxy disks comparable to the Milky Way is
typically in the range $5 - 10 \, \rm M_{\odot} \, \rm yr^{-1}.$
Non--axisymmetric instabilities,
such as spiral density waves, are the underlying trigger of star formation,
most of which occurs in the spiral arms.  These may be driven by a central
bar, by the tidal interaction with a companion galaxy, or could even erupt
spontaneously as a consequence of the amplification of stochastic noise.

Much higher star formation rates are associated with starburst galaxies.  Here
the SFR may be one or even two orders of magnitude higher, per unit mass in
stars.  Compelling obervational evidence suggests that many starbursts, and all
of the extreme starbursts, are driven by galaxy mergers.

The early history of star formation in the Milky Way is inferred to have been
relatively quiescent, not differing by more than a factor of 2--3 from the
present day SFR \ref\twa\ref\bar\ref\nohs$[\twa,\bar,\nohs.]$  This applies to
our galactic disk.  One can infer star
formation rate histories for nearby spiral galaxies, and similar results are
found.  The SFR in the late--type spirals (Sd) actually increases slowly with
time, whereas the SFR in Sa's and Sb's  decreases \ref\gal$[\gal].$

A much higher star formation rate per unit mass is inferred for spheroidal
stellar populations.  The lack of young stars requires all star formation to
have terminated at least $6 \rm Gyr$ ago.  Population synthesis modelling
requires
the bulk of the star formation to have occurred in $1 \rm Gyr$ or less.
Therefore
in spheroids, the SFR was an order of magnitude or more higher than in disks.

{\it 3.2. The Primordial IMF}

One might expect the IMF to be different for extremely
metal--poor stars, if only because many of the processes involved in
fragmentation and star formation are sensitive to metallicity.  Our best
indicator of the primordial IMF comes from examining heavy element abundance
ratios in metal--poor stars.  From these studies, one can make crude inferences
about the IMF of the precursor stars that synthesized the
metals \ref\s$[\s].$
 The odd--even
pattern of abundances seen in extreme metal--poor stars is identical to that in
star formation at the present epoch.  The $r$ and $s$ process sites are
believed to
be massive stars. Hence massive stars $(10-100 \rm M_\odot)$ were present in
the
first stellar population.  Low mass stars were also present. Stars of
$\sim 1 \rm M_\odot$ are found with [Fe/H] $ < -4.$ These stars  are
sufficiently metal--poor
that the observed enrichment should have occurred during the first generation
of
star formation.

Theoretical considerations of the fragmentation of primordial clouds result in
predictions of minimum fragment masses that differ little from similar
predictions for clouds of solar
abundance.  Opacity--limited fragmentation proceeds as follows.  A collapsing
cloud is initially transparent to radiation, and cooling regulates the collapse
to be approximately isothermal at temperature $T.$  The Jeans mass,
proportional to
$T^{3\over2} \rho^{-{1\over2}},$ therefore decreases until the density $\rho$
is sufficiently high that the optical depth across a fragment is appreciable.
The
ensuing collapse is nearly adiabatic, so that the Jeans mass is proportional
to $\rho^{{3\over 2} (\gamma - {4\over 3})},$ with $\gamma \approx {5\over 3}.$
 The minimum Jeans scale is about $10^{-3} \rm M_{\odot},$ and is only weakly
dependent  on metallicity.  For a cloud of solar abundance, grain cooling is
dominant, whereas for a primordial cloud, molecular hydrogen formation and
dissociation control the cooling and fragment mass evolution.  In both
situations, the effects of finite size of the parent cloud increase the
minimum fragment
mass scales as fragments can shadow one another and thereby enhance the
effective  opacity.

 Other physical effects that are difficult to model but
that nevertheless are important include fragment collisions, fragment mergers
and
accretion of diffuse gas by fragments.  The turbulent velocity field induced by
asymmetric collapse and by
feedback from forming stars will help drive fragment interactions.  The
general sense of these modifications of the naive, spherically symmetric
treatment of opacity--limited fragmentation is to drive the minimum fragment
mass up to at least $0.01 \rm M_{\odot},$ and perhaps to $0.1 \rm M_{\odot}.$
This
could therefore account  for the paucity of brown
dwarfs in conventional star formation.  With regard to primordial star
formation, the prospects for BDM being mostly in the form of brown
dwarfs are evidently dim.  There is no compelling reason that primordial
conditions would systematically favor domination by fragments of mass below
$0.1 \rm M_{\odot}.$

Accretion onto protostellar cores is parametrized, in a simple spherically
symmetric situation, by the accretion rate $\sim {{\Delta V^3} / G},$ where
$\Delta V$ represents an effective throttle velocity at which inflow occurs.
This might be the sound velocity in a quiescent cloud, the turbulent velocity,
or the Alfven velocity if magnetic pressure dominates the thermal pressure.  If
one
could imagine an unusually quiescent enviroment, with an accretion rate as low
as $\sim 10^{-9} \rm M_{\odot} yr^{-1},$ it is possible to delay hydrogen
ignition and construct brown dwarfs
of mass $0.1$ or even $0.2 \rm M_{\odot}$  \ref\salp$[\salp].$ More normal
accretion rates are in the
range $10^{-5} - 10^{-3} \rm M_{\odot} yr^{-1}.$  These lead, for low mass
cores, to
conventional brown dwarfs, of mass below $0.08 \rm  M_{\odot}$ for solar and
$0.09\rm  M_{\odot}$
for primordial composition.  It is possible that in a turbulent
cloud, where $\Delta V$ is enhanced, as well as in a primordial cloud, where
inefficient cooling guarantees a high sound speed, the protostellar accretion
rates are large.  This would provide a possible theoretical justification for a
top--heavy IMF in these environments.

Halo BDM could conceivably consist of stellar relics if the primordial IMF had
very few solar mass stars. Indeed that the primordial IMF was top--heavy is at
least as likely
as the
bottom--heavy option. Several arguments may be adduced to support this
possibility
\ref\Silk$[{\Silk}].$
The low dispersion found in the alpha--nuclei relative to
iron
$[{\gus}],$
 compared to the large dispersion in [Fe/H] for disk stars,
suggests that both
the $\alpha$--nuclei and Fe were mostly produced by massive stars, in contrast
to the current epoch IMF that generates Fe from low mass Type I supernovae and
$\alpha$--nuclei from massive stars.  The conventional interpretation that only
at $\rm [Fe/H]<-1$ is one dominated by massive star--synthesized alpha--nuclei
is probably not
tenable in view of recent data
\ref\bes
$[{\bes}],$
which reveals a gradual trend of decreasing
$\alpha$--nuclei with increasing iron abundance.  The enhancement with
decreasing galactic radius
$[{\gus}]$
of
the alpha-nuclei abundance relative to Fe/H
suggests that the primordial IMF in the inner galaxy was systematically
top--heavy relative to the
solar neighborhood.

This latter possibility is also suggested by the analogy between galaxy
formation and starbursts.  The elevated star formation rate inferred when the
old disk and spheroid formed is similar to that encountered in starburst
galaxies. The physical mechanism, involving satellite mergers, is common to
models of both starbursts and galaxy formation.  Modelling of starbursts
suggests that a top--heavy IMF is required to account for the observed
luminosity, given the available gas supply and a plausible star formation
efficiency.

One might expect the same situation to have applied when the inner
galaxy formed and the bulk of the heavy elements seen in the disk were
synthesized.  Stellar remnants provide an attractive source of mass to
account for the rotation curve in terms of a boosted contribution from the
inner disk
\ref\lar$[{\lar}].$
It has also been suggested
\ref\san$[{\san}]$
that without a top--heavy IMF at early
epochs one would have exhausted the supply of interstellar gas by the present
epoch. A top--heavy IMF in the inner galaxy may be required at the present
epoch to account for the observed ionizing photon flux $[{\mez}].$
Overproduction of $^3$He is avoided with an early IMF that has fewer, by a
factor of 2--3, low mass stars than the present--day IMF. One cannot overdo
this, otherwise there would be excessive astration of $^2$H.

If a primordial top--heavy IMF is held responsible for disk and spheroid
formation, it is evidently possible to flatten the IMF still further, or
even truncate it below $\sim 2 M_\odot$, in order to account for halo BDM.
The dominant
component is most likely to be white dwarfs, since their stellar precursors
$(<10
M_\odot)$ produce relatively little light or nucleosynthetic contamination
compared to more masssive stars.  This requires fine--tuning of the IMF. For
example, with an IMF only spanning
a range of $\sim2-8 M_\odot$,
one can avoid excessive CN production, since primordial
stars with very low abundance $(Z<10^{-4}Z_\odot)$ do not undergo helium
flashes and ensuing
dredge--up of CN--cycle processed material \ref\chit\ref\fuji$[\chit,\fuji].$
$^4He$ production and $^2H$
destruction cause potential difficulties. Both uncertainties in the primordial
abundances and the likelihood of considerable gas recycling  offer
considerable leeway. It is quite possible that the gas may reside in the halo
or outer disk in the form of cold clouds, or else be
ejected into the intergalactic medium when the outer BDM halo forms
\ref\ryu$[{\ryu}].$

 {\it 3.3 What Could the (Dark) Matter Be?}

If the dark matter is baryonic, it makes sense to consider the most reasonable
forms that it could take.  These are, in order, of decreasing plausibility:

\item {\it a.}    Stellar mass objects, from $10^{-3} M_{\odot}$
to $10^{3}M_{\odot}.$  These could be brown dwarfs
$(10^{-3}$ to $\sim 0.08  M_{\odot}),$ white dwarfs
$(0.4 - 1.4 M_{\odot}),$ neutron stars $(0.4 - 2 M_{\odot}),$  stellar relic
black holes from ordinary massive stars $(\sim 2 - 10 M_{\odot}),$
or black holes that formed from supermassive stars $(\sim 100 - 10^4
M_{\odot}).$

\item {\it b.}  Diffuse dense clouds of cold hydrogen,

\item {\it c.}  Exotica, including primordial black holes and nuggets of
strange
matter.

Unfortunately, as we have seen, theory is a poor guide. The physical conditions
in primordial clouds undoubtedly differ from present--day star--forming clouds.
There were no heavy elements, no dust, and, most likely, no significant
magnetic fields. However we have so sparse an understanding of how the
present--day IMF arises that it is not even possible to infer the sign of any
deviation in the primordial IMF from that observed locally. We cannot predict
whether the primordial IMF should be biased towards massive or low mass stars.
Biased it must be, however, in order to produce sufficiently dark matter.

Black holes of mass larger than $10^4 M_{\odot}$ have been recently excluded as
a halo dark matter candidate, since otherwise globular clusters
\ref\moo$[\moo]$
and nearby dwarf spheroidal galaxies \ref\rixl$[\rixl]$
would be disrupted.
Stellar mass objects are preferred, as doing the least injustice to our
expectations of what halo dark matter might be.  One can distinguish between
the various options for stellar mass objects on astrophysical grounds.

{\it 3.3.2   Brown Dwarfs}

Not a single brown dwarf is known to exist. Intensive searches
for low mass stellar companions of nearby stars by spectroscopy (to detect
motions of $\sim 20 \rm \, m\, s^{-1}$) and photometry (to measure shifts of
$\sim 0.001$
arc-sec/yr) have failed to reveal any candidates below $0.08  \rm M_{\odot}$
\ref\burl\ref\bess$[\burl,\bess].$
Searches of binaries have failed to find companions of later spectral types
than $M6.$
Spectroscopy of candidate brown dwarfs that are in the Pleiades, chosen from
their location in the H--R diagram, failed to find lithium absorption lines
\ref\marb$[\marb].$  These
stars cannot therefore be brown dwarfs.
These failures to find brown dwarfs have not dissuaded theorists from proposing
brown dwarfs as BDM candidates.

The population II IMF appears to  be steep, rising to the detection limit of
about $0.15\rm M_\odot.$  However the density profile follows that
of a de Vaucouleurs law \ref\ric$[\ric].$
A halo of brown dwarfs is
directly detectable if the IMF of the halo is an extrapolation of the IMF
observed in Population II stars.  This is because even though one would need to
extrapolate the IMF to very low masses, $0.01 - 0.001 M_{\odot},$ given any
reasonable slopes, the brown dwarfs just below the main sequence limit are
still
sufficiently luminous after a Hubble time has elapsed to be detectable via deep
star counts in the near-infrared.  Recent surveys at $2.2 \mu$  suggest that a
brown dwarf halo could not be a simple extrapolation of the IMF of Population
II stars \ref\huh$[\huh].$  However, a strong upturn in the IMF below the main
sequence limit of
$0.09  \rm M_{\odot}$ for primordial brown dwarfs or  $\sim \rm  0.08
M_{\odot}$ for
metal-rich brown dwarfs would not be detectable in the deep counts.  Only the
gravitational microlensing surveys provide an unambiguous means of searching
for these low mass halo objects.

The pro--brown dwarf arguments are the following.  Brown dwarfs presumably form
in considerable numbers, since a fragmenting interstellar cloud is unaware of
the minimum mass for hydrogen burning.  Cooling flows in galaxy clusters
are inferred to undergo mass deposition at a rate of up to $\sim 300  \rm
M_\odot
\rm yr^{-1}$, and this mass flux cannot end up in stars with a solar
neighborhood
initial mass function (IMF).  Some evidence of star formation is seen, however,
for example in the galaxy NGC 1275 at the centre of the Perseus cluster cooling
flow \ref\ricc$[\ricc].$ This suggests that the IMF formed in cooling flows may
be bottom--heavy,
or steeper than the local IMF.

Evidence \ref\white$[{\white}]$
of
$\sim 10^{11}-10^{12} \rm M_\odot$ in cold gas in the cores of cooling flow
clusters, based on
modelling the x-ray spectrum below 1 keV, may, if confirmed, remove much of the
motivation for invoking predominantly low mass star formation in cooling flows.
 Tentative support for a bottom--heavy IMF comes from a study of several
globular
cluster luminosity functions.  In the mass range $\sim 0.2-1 \rm M_\odot$,
despite
large incompleteness corrections, a significantly steeper IMF is found for a
few
 globular clusters, several of which have long core relaxation
time--scales \ref\richter$[{\richter}].$
However, the metal abundance in these systems is an order of magnitude lower
than that in cooling
flows, $[Fe/H]\approx -0.5$.

{\it 3.3.3 Halo White Dwarfs}

White dwarf mergers are believed to result in Type I supernovae.  These are
luminous and catastrophic events that are powered by the ejection of about
$0.6  \rm \, M_{\odot}$ of radioactive nickel that decays into iron.  A dark
halo
would be
detectable were it to generate Type I supernovae at a rate expected for the
corresponding number of white dwarfs.  However there is some reason to believe
that Type I supernovae are subluminous in old stellar populations
\ref\filet$[\filet].$

A white dwarf halo requires extreme fine--tuning of the primordial IMF. One has
to obtain a mass--to--light ratio of $\simgt 2000$ in the V--band, as inferred
from observations along the minor axes of edge--on spirals $[\skr].$ One has
also to avoid contamination by ejecta from supernovae. The allowed mass range
of the precursor population is $4-6\rm\, M_\odot$ if the stars form in a burst
that lasts 2 Gyr,
and  $2-8\rm\, M_\odot$ if the burst lasts 1 Gyr $[\ryu].$

If one removes the assumption that the massive star ejecta are recycled in the
disk, the constraints on the upper end of the IMF can be relaxed. For example,
the gaseous ejecta from the halo could be ejected into the intergalactic medium
by supernova--driven winds, which would then give a source for the intracluster
iron detected in x--ray observations of rich clusters. The abundances of other
elements in the intracluster gas, especially oxygen, will soon be available
from ASCA observations, and should help clarify the nature of the parent star
population that must have contaminated the intracluster medium early in
galactic history.

 A massive--star origin would result in enhanced oxygen to iron by a factor of
3 or so, as seen for old Population II stars. This would be true for an IMF
truncated at the lower end \ref\wans$[\wans].$
However a precisely fine--tuned IMF, greatly but not necessarily completely
suppressed at both lower and upper ends in order to produce white dwarf halos,
would result in a more normal abundance of oxygen relative to iron.
Alternatively, the halo gas left over from forming the stellar relic BDM,
amounting to as much as 70 percent for a typical return fraction appropriate to
a top--heavy IMF, could have condensed into cold gas clouds that remain in the
halo, as discussed in the next section.

A halo of white dwarfs, with minor components of neutron stars, black holes,
and
even solar mass stars, as predicted by a top--heavy IMF, is potentially
observable via several experiments.  If the halo formed less than $\sim 15$ Gyr
ago, the white dwarfs are sufficiently luminous $(L>10^{-6} \rm L_\odot)$ that
the
nearest ones are observable: for example, a frequency of $\simgt$ 1/sq deg to
$m_I<22$ is predicted
\ref\tam$[{\tam}].$

Perhaps the most dramatic consequence of a white dwarf halo
stems from binary mergers.
 Mergers of close
white dwarf pairs formed by tidal capture in the protoclusters where they were
formed could produce neutron stars.
The
required production mechanism must form a substantial number of high galactic
latitude
pulsars, seen at a distance above the
galactic plane  $\simgt 1 \rm \, kpc,$ and
generates pulsar velocities of $\simlt 1000 \, \rm km\,s^{-1}.$
Neutron stars formed via
mergers are
plausible candidates for
gamma--ray bursters \ref\eic$[\eic].$  The existence of high velocity pulsars
in the halo is suggested  by the observational
 data. The possibility of their being gamma--ray burst
progenitors is to a large degree independent of the
theoretical model.

  {\it  3.3.4 Diffuse Gas Clouds}

The most conservative of assumptions for the nature of dark matter is that it
is in the form of diffuse gas.  In galaxy halos, this at first sight seems to
be completely untenable.  Gas at the virial temperature of the Milky Way,
$\sim 2 \times 10^6 \rm K,$ would prolifically emit soft x--rays.  The diffuse
x--ray
background allows an x--ray emission measure of at most $0.01 \,\rm cm^{-6}\,
pc,$ corresponding to a halo density at $\sim 10\rm \, kpc$ (the dark halo core
radius) of
$\sim 10^{-3}\rm \, cm^{-3},$ and therefore to a mass of $\sim 10^8  \rm
M_{\odot}.$  With
a density profile $\rho \propto r^{-2},$ the diffuse gas mass is only
$10^{-3}$ of that required for the dark halo.

 However several observations
suggest
that one ought to reexamine the diffuse gas constraints more carefully.  The
deepest x--ray observations of galaxy clusters indicate that considerable
amounts of hot gas may be outside the cluster core.  In several clusters, the
gas mass
amounts to more than $50$ percent of the total mass at 3 or 4 Abell radii
\ref\bri\ref\wat\ref\eyl$[\bri,\wat,\eyl].$
Moreover in the inner cores, where cooling flows are inferred from the x--ray
surface brightness profiles, there are indications of x-ray self absorption
intrinsic to the cluster.  These are best interpreted in terms of
$\simgt 10^{11} \rm  M_{\odot}$ of cold  gas, inferred to be in clouds with a
covering
factor of order unity across the cluster core \ref\whif$[\whif].$  High
redshift observations of
damped Lyman alpha clouds indicate that if the trend observed at
$z \simgt 3,$ where one measures a mass fraction in hydrogen that is roughly
equal to that seen in stars at $z=0$ \ref\wolf$[\wolf],$ continues to $z\sim 5$
one may be seeing more cold gas in the form of HI than is in stars at low
redshift.

With regard to our own galaxy, there may be several times more molecular gas
than atomic gas in the disk at a galactocentric distance of $\sim 10 \rm kpc$
\ref\leqa$[\leqa].$
Indeed, molecular gas complexes, excited by HII regions, have been discovered
as far out as $\sim 28 \rm kpc$ \ref\deg$[\deg].$ There may well be far more
colder $H_2$ present in the
outer disk, without accompanying HII regions, than has hitherto been
undetected.  If this trend  were to continue to the outermost disk, at $\simgt
30 \rm kpc,$ one
might need to revise the consensus view that the mass in cold gas does not
contribute significantly to the rotation velocity.

The remarkable case of DDO 154 provides strong testimony for the view that dark
matter normally associated with halos may exist at least in part in the form of
hitherto
undetected cold gas clouds.  This dwarf galaxy has one of the best-studied
rotation curves, that extends to at least $15$ disk scale--lengths.  Outside 2
scale--lengths the observed star distribution provides a negligible
contribution to the
rotation velocity. The HI column density scales as $N_{HI} \propto r^{-1}$ to
the limit  where
the rotation curve can be traced.  It exactly parallels the dark matter surface
density inferred from the rotation curve, and contributes about 10 percent to
the required total surface density.  It is tempting to infer that the 90
percent shortfall of dark matter is in baryonic form, either $H_2$ or optically
thick HI.  It might be physically associated with the observed HI in the disk,
or else constitute a flattened halo in which the observed HI clouds are
embedded.

Hiding a population of cold clouds from detection is possible if the clouds
are sufficiently compact so as to only rarely  collide.  This same condition
also guarantees that the cloud surface covering factor is low.  It would then
be difficult to observe the clouds, either in absorption towards quasars or in
emission.  There are two difficulties.  Clouds passing through the disk would
be exposed to the local ionizing radiation field within HII regions and
possibly be visible.  The overriding question is why such clouds avoid forming
stars during a Hubble time.  Stabilizing the clouds is possible if pressure
support can be maintained.  One would need warm cloud cores.  It might be
possible to achieve this with a modest amount of star formation.  Primordial
abundances in the cores would also result in higher temperatures.

 {\it 3.3.5  Exotica}

Baryon dark matter could consist of massive black holes.  As noted above, the
upper limit on black hole mass is about $10^4 \rm M_{\odot}.$  Precursor
supermassive stars in the mass range $100 - 1000  \rm M_{\odot}$ implode to
form black holes without injecting substantial amounts of enriched material.
However, during the precollapse helium-burning phase, there is extensive
radiatively-driven mass loss, and considerable amounts of helium are shed.
To avoid a discrepancy with primordial nucleosynthesis, one would have to
store the ejecta in cold dense clouds that remain in the halo or outer disk.
These clouds cannot participate in spheroid and disk star formation, and are
not otherwise strongly constrained, as described in the previous section.

Nuggets of strange matter, relics of the quark--hadron phase transition, have
been proposed as  a possible form for dark matter. Such objects may be stable,
in certain quark models. However, quark nuggets are likely to have evaporated
prior to the nucleosynthesis epoch.

The smallest stable objects that might be BDM candidates have masses that can
be estimated as follows \ref\derj$[\derj].$  These would be made  of hydrogen.
For a density of
solid $H_2$ of about $0.1 \rm g\, cm^{-3},$ such ``snowballs" are
gravitationally bound
at a temperature of say $30\rm  K,$ corresponding to the CMB temperature at
$z=10,$ if the typical mass exceeds
$$ M \simgt 10^{-8} \left( T \over 30 \rm K\right)^{3\over 2}
\left({0.1\rm g\, cm^{-3}} \over \rho\right)^{1\over 2}\rm  M_{\odot}$$
Thus the mass range $10^{-8} \rm M_{\odot}$ to $10^{-3}\rm M_{\odot}$ is the
possible
range spanned by dark matter snowballs.  The central pressure is
sufficiently high that if the mass exceeds $\sim 10^{-3} \rm M_{\odot},$
degeneracy
is important.  More massive objects have higher central density and are
smaller.
 They  continue to contract, although at a small rate, and are referred to  as
brown dwarfs.

In summary, I conclude that star formation is a messy problem in nonlinear
physics with depressingly many degrees of freedom. These include cloud
ionization, metallicity, magnetic field strength, angular momentum, dust grain
properties, and possible feedback from forming stars. At least, we can predict
the mass of a star, to within an order of magnitude! I have argued that
phenomenological arguments provide a useful guide. Unfortunately, a mastery of
star formation is critical for  understanding the nature of baryonic dark
matter.  One needs either to prevent star formation from occurring, as is the
case if BDM  consists of cold clouds or brown dwarfs, or else to fine--tune it,
as must be done if BDM is in the form of white dwarfs or black holes.

 {\bf 4. COSMOGONIC IMPLICATIONS}

 {\it 4.1 Galaxy Morphology}

The primordial star formation rate is the key to understanding galaxy
morphology.  The high specific star formation rate inferred during formation of
the spheroidal component of a galaxy guarantees that some massive dense stellar
subsystems form early in the collapse.  These sink deep into the potential well
via dynamical friction against the lower density stellar systems that are the
prevalent component.  The dense star clouds efficiently transfer angular
momentum as they spiral into the central regions of the galaxy.  In this way,
an elliptical galaxy develops that is supported by random stellar motions
rather than by systematic rotation.  The star formation is completed within 1
or 2 Gyr.

In contrast, a disk forms slowly, over several Gyr.  The low star formation
rate means that the system stays gas-rich.  Dissipative cooling controls the
rate at which the angular momentum-conserving contraction occurs.  Eventually,
rotational support halts the collapse process, when the disk has formed.  The
role of a dark halo is to provide an additional source of mass-collapsing
matter against which the gaseous, star-forming component exerts a torque and
thereby transfers angular momentum.

Evidently, star formation plays a crucial role in determining the various types
of galaxies.  Galaxy halos are likely to have some BDM, and perhaps to be
predominantly BDM.  The formation of baryonic dark matter, since it is closely
coupled to early star formation, is evidently inseparable from the galaxy
morphology issue.  An explanation for why spirals predominate in low density
regions and ellipticals in dense cluster cores is likely to be related to the
problem of BDM.

{\it 4.2   Large-Scale Structure}

I have hitherto assumed that halo dark matter consists of BDM.  This is
equivalent to asserting that
$\Omega_{BDM} = 0.03 - 0.07,$ in accordance with the nucleosynthesis prediction
$\Omega_B = 0.015 (\pm 0.005)h^{-2}.$
However, there is reason to doubt the error bounds on the nucleosynthesis
limit.  If these are sufficiently relaxed, one is then drawn to consider the
case of an open cosmology with $\Omega = 0.1 - 0.2,$ in which all of the dark
matter is BDM.

Could one go the additional step and consider $\Omega_{BDM} \approx 1$~?  This
would
grossly violate the nucleosynthesis limits even in non-standard models of
inhomogeneous light element production.  Such a model almost certainly produces
excessive cosmic microwave background fluctuations.  Certainly with
inflationary
initial conditions, a primary motivation for adopting  $\Omega = 1,$ one has
approximately scale-invariant, adiabatic primordial density fluctuations.
These are a disaster for ${\delta T} / T;$ nor is ${\delta T} / T$
suppressed by reionization on the largest angular scales.

However, the $\Omega = 0.1 - 0.2$ cosmology is phenomenologically attractive.
It makes the simplest of assumptions:  ``what you see is what you get.''  We
see baryons, and on scales $< 20 \rm Mpc, $ where the observations are most
reliable, we measure $\Omega \sim 0.1.$
There is a heavy price to pay for the simplicity.  One has to drop inflation,
at least in its generic incarnation, and one has to abandon the hypothesis of
primordial scale--invariant curvature fluctuations.

The result is a model that is ugly but simple.  A low $\Omega$  universe,
containing only baryons, must be seeded by primordial isocurvature
fluctations.  These are equivalent to primordial spatial variations in the
specific entropy or in the baryon number.  There is no accepted theory for the
origin of such fluctuations.  However, one might anticipate that some models of
baryogenesis, for which there is not a universally accepted theory, and which
provide ${n_B} \over {n_{\gamma}},$ are also capable of producing
$\Delta \left( {n_B} \over {n_{\gamma}}\right).$  Indeed, there are such
models in the literature \ref\dols\ref\yoks $[\dols,\yoks].$  However there are
essentially no predictions for the
fluctuation spectrum, which accordingly is treated phenomenologically, as a
power-law of arbitrary slope and normalization.

Primordial entropy perturbations $\delta s$ are defined as perturbations in the
number of photons per baryon, so that
$$\delta s = \delta \left( {T^3}\over n\right),\ \ {\rm whence}\ \  {\delta
s\over s}= {3\over 4} {{\delta
\rho_\gamma}\over{\rho_{\gamma}}} - {{\delta \rho_B}\over {\rho_B}}.$$
Here, $\delta \rho_{\gamma}$ is the perturbation in radiation density and
$\delta \rho_B$ is the perturbation in baryon density.  Requiring that there be
no net curvature perturbation, the isocurvature mode being orthogonal to
the adiabatic or curvature mode, then leads one to write
$$\delta \rho_{\gamma} + \delta \rho_B = 0.$$
In the late time, matter-dominated limit, one obtains
${{\delta T}\over T} = {1 \over 3} \delta s.$
This is valid on scales larger than the horizon at
last  scattering, and shows that one can map out the intrinsic entropy
fluctuations on sufficiently large angular scales (greater than a few degrees).

In the absence of a predicted spectrum, one adopts a power-law form for the
primordial entropy fluctuations, with power spectrum $P(k)\equiv \left|
\delta_k\right|^2 \propto k^n,$ where $\delta_k$ is the Fourier amplitude,
$${\delta \rho\over \rho} = \int \delta_k\, exp (i{\bf k.x})\, d^3k.$$

The rms fluctuations $\langle ({\delta \rho /\rho})^2\rangle$ are equal to
$\left|\delta_k \right|^2.$ An empirical fit to the COBE DMR data over
spherical harmonics $(l =2 - 10)$ finds $n \approx 1.1$ but with large
uncertainty, $\Delta n \approx \pm 0.5.$  Comparison with the Tenerife data
$(l \approx 18)$ suggests that $ n \approx 1.5,$ as does analysis of the second
year DMR data.  If confirmed, this would favour a non--inflationary primordial
fluctuation spectrum as expected in a low $\Omega$ universe.

 {\it 4.3 Primordial Density Fluctuation Power Spectrum}

An empirical fit to the matter fluctuations can be performed using the power
spectrum derived from various redshift surveys.  Over scales of $10 - 50 \rm
Mpc,$
the linear regime of power is effectively probed, albeit with uncertainties
that depend on the inevitable distortions involved in transforming from
redshift to three-dimensional space.  An empirical fit requires $n \approx -1,$
with an uncertainity of about $\Delta n \approx \pm 0.5.$  The more negative
values of $n$ result in excessive CMB temperature fluctuations on scales of
order $10$ degrees, where reionization is ineffective.  A compromise value is
$n \approx -0.5$ for the primordial power law index.  In terms of the invariant
mass $M$ associated with comoving wavenumber $k,$ the corresponding mass
spectrum is
${{\delta \rho}/ \rho} \propto M^{-{{n+3}\over 6}} \propto M^{-0.4}$ for $n=-
0.5.$  Hence in contrast to the scale-invariant inflationary spectrum
$(n\approx 1),$ which is only logarithmically divergent, with $n_{eff}
\approx n - 4,$ on scales smaller than that of the horizon at matter-radiation
equality, roughly a galactic mass, the isocurvature spectrum is strongly
divergent towards high redshift.

  Early formation of small galaxies is
 inevitable in this model.  Star formation, and the associated
supernovae, must result in production of an ionizing photon flux that is
capable of at least partially reionizing the intergalactic medium.  Even with a
small efficiency of ionizing photon production, recoupling of the CMB is likely
to be almost inevitable at $z \simgt 100.$  This has two notable effects.
Radiation drag inhibits growth of matter fluctuations on sub-horizon scales.
Rescattering of the CMB smooths out the associated temperature fluctuations.

To produce the large-scale structure, as characterized by the correlation
amplitude on $10 \rm Mpc,$ the suppressed growth implies that one needs a
larger initial amplitude for the primordial fluctuations than would be the case
were
early reionization $(z > 100)$ not to have occurred.  This has interesting
consequences for the generation of large-scale peculiar velocity fields.  The
baryonic dark matter power spectrum has a generic large-scale peak that
corresponds to the maximum Jeans mass scale.  This is approximately equal to
$110 \left(
{0.1} / {\Omega h^2} \right)\rm Mpc.$

The sound speed prior to recombination is
$$c_s = \left ( dp \over d\rho \right )^{1\over 2} =
\left( {{dp_r}\over{d(\rho_m + \rho_r)}}\right)^{1\over 2}
 = {c \over {\sqrt 3}} \left( 1 + {3\over 4} {\rho_m
\over \rho_r}\right)^{-{1\over 2}} \propto (1 + z)^{1\over 2}$$
at $\rho_m > \rho_r.$
The Jeans length $l_J \sim c_st \propto (1 + z)^{-1},$ and therefore the
comoving Jeans length is constant.  After recombination, the temperature
abruptly drops to $3000 \rm \, K$ and the sound speed correspondingly declines.
 The
Jeans mass prior to recombination was $2 \times 10^{18} \left ({0.1 \over
{\Omega h^2}}\right)^2 \rm M_{\odot};$ after recombination, it drops to $ 5
\times
10^5 \left ({0.1 \over {\Omega h^2}}\right )^{1\over 2} \rm M_{\odot}.$
Since the sound speed prior to recombination is about
$10^4$ times larger than that after recombination, any pressure fluctuations
propagating as sound waves are greatly amplified over scales much larger than
the post-recombination Jeans length.

One consequence is the occurrence of dramatic oscillations in the matter
transfer function (Figure 1) that are eventually quenched by radiation drag, in
a universe where reionization occurs at $z<1000.$  It is unclear whether the
corresponding oscillations in the galaxy correlation function, the  amplitude
of which depends sensitively on the model for small--scale nonlinearity, would
be observable. A large--scale coherent velocity field is another consequence of
this model \ref\peeb$[\peeb].$ Large--scale bulk flows are more directly
computable, being insensitive to any non--linear corrections for the range of
$n$ of interest,
and measurable.
The velocity correlation function is shown in Figure 2. The large--scale matter
distribution is normalised to give unit variance in mass fluctuations averaged
over a sphere of radius $8h^{-1}\, \rm Mpc,$ where the luminous galaxy counts
have unit variance.

 {\it 4.4 ${{\delta T}\over T}$ on Intermediate and Small Angular Scales}

The initial expectation for ${{\delta T} \over T}$ in a low $\Omega$ universe
is that the Jeans mass peak would have a substantial effect.  On scales below
the horizon at recombination, corresponding to several hundred Mpc, gravity
would amplify the primordial entropy fluctuations.  Associated adiabatic
fluctuations are generated by gravity-induced velocity fields.  This should
lead to ${{\delta T} \over T} \sim {\upsilon \over c} \sim {L \over t} {{\delta
\rho} \over \rho}.$   However the first-order Doppler
fluctuations are erased by rescattering of the CMB photons.  The probability
of rescattering is
$$\int\limits^{to}_t\ n_e \sigma_{T} cdt = 0.04 h \Omega_B \Omega^{-
{1\over2}}_0 (1 + z)^{3\over 2}.$$
With $\Omega_B \sim \Omega_o \sim 0.1,$ primary fluctuations are erased if
reionization occurs at $z \simgt 50,$ over angular scales of up to $\sim 10$
degrees.

However, temperature fluctuations are regenerated on the last scattering
surface.  Only in second order, ${{\delta T} \over T} \sim ({v \over c}) (
{{\delta \rho} \over \rho}),$ do the fluctations add in quadrature, the
first-order fluctuations $(\sim {v \over c})$ self-cancelling.  While any
surviving
first-order fluctuations would be on degree scales, and correspond to the
primary Doppler peaks, the second order, regenerated fluctuations are on
arc-minute scales.  These are a unique signature of BDM on these scales,
since the
primary last scattering surface has a thickness of about 5 arc-minutes, and in
the canonical CDM model, one expects no primary fluctuations on smaller scales.

The best current limit on ${{\delta T} \over T}$ over small angular scales is
that from the Australia Telescope Compact Array.  Over a beam of $0'.9,$ the
rms temperature fluctuations are less than $9 \times 10^{-6}.$  This allows a
small area of BDM parameter space, with $\Omega_B \sim 0.1,$ $n \sim -0.5,$ and
$h \sim 0.8$ \ref\hus$[\hus].$ With the Hubble constant as low as $h=0.5,$
excessive
fluctuations are generated on arcminute scales.

 {\it 4.5 The Compton y Constraint}

A significant spectral distortion of the CMB blackbody arises if reionization
occurs very early.  Compton scattering not only erases angular fluctuations,
but transfers energy from the hotter electrons to the CMB photons,
${{\Delta h \nu}\over {h\nu}} \sim {{kT} \over {m_e c^2}}.$  The resulting
distortion is a simple function of the Compton  $y$ parameter, defined by
$$y = \int\limits^{t_o}_{t_{reion}}\  n_e\ \sigma_{T}\ cdt (
{{kT}\over {m_e c^2}}) \approx 6.4 \times 10^{-8} h \Omega_B T_4 z^{3\over
2}_{reion}  ,$$
where $T_4 \equiv {T \over 10^4}\rm K $ is the temperature of the intergalactic
gas. The COBE FIRAS experiment sets an upper limit,
$$y < 2.5 \times 10^{-5}.$$
This is sufficient to exclude models in which reionization occurred at
$z \simgt 800,$ with $h \sim 0.8,$ $\Omega_B \sim \Omega_0\sim 0.1$
\ref\tegs$[\tegs].$

 {\it 4.6 A BDM Scenario}

Consider the following model for a cosmology dominated by BDM.  Take $\Omega
\sim 0.1 \sim \Omega_B.$  This alone requires early structure formation, even
clusters of galaxies forming at $z \simgt 10.$  Galaxies form much earlier.
With primordial entropy fluctuations allowed as possible seeds, adiabatic
fluctuations being observationally excluded, one infers a linear fluctuation
distribution described by
$${{\delta \rho}\over \rho} \propto M^{{-{1\over2}}{-{n\over 6}}} t^{2\over 3},
\ \ z \simgt { {\Omega^{- 1}-1}},\ \  0 \simgt n\simgt -{0.5}.$$
Nonlinearity occurs on the Jeans mass scale, $\sim 10^6 M_{\odot},$ as early as
$z \sim 1100.$

What happens next is pure speculation.  One scenario is the following.  The
primordial clouds, of mass comparable to globular star clusters, collapse, and
fragment into stars by $z \sim 500.$  Ionizing photons from the first massive
stars ensure that Compton drag forces will initially inhibit further gas
collapse
and star formation.  However once $z \simgt 200,$ the cloud contraction is
sufficiently shorter than the Hubble time that star formation resumes.
Supernovae drive gas outflows that will soon disrupt star formation in the low
mass clouds.  Only later, by $z \sim 30,$ when sufficiently massive potential
wells have developed that can efficiently retain the ejection from
supernova--driven winds, will galaxy formation begin in earnest.  The
baryonic dark matter
consists in part of the compact remnants of early massive star formation, and
also, at least in the intergalactic medium, of diffuse gas.

 {\bf 5. CONCLUSIONS}

The BDM hypothesis provides a reasonably complete description both of dark
halos and of dark matter that is more broadly distributed.  It leads to five
unique predictions.  Three of these are related to CMB temperature
fluctuations.

\item {a.} Secondary fluctuations at a level
${{\delta T} \over T}\sim 10^{-5}$ are
predicted on arc-minute scales because of the early reionization that the BDM
model requires in order to erase the primary Doppler peaks.

\item {b.}
Secondary Doppler peaks $\delta T/ T \sim 10^{-5}$ are generated on degree
scales. Their location depends on $\Omega.$

\item {c.}
On large angular scales, $\simgt 10$ degrees, curvature effects dominate the
predicted temperature  anisotropies.  In addition to the usual
Sachs-Wolfe anisotropies, ${{\delta T}\over T} = {1\over 3} \phi_{LS},$ from
the last scattering surface, there is the integrated effect of time-varying
potentials along the line-of-sight, ${{\delta T} \over T} = \int {{\partial
\phi}
\over {\partial t}} dt.$  The curvature scale
$({c \over H_0}) (1 - \Omega_0)^{-{1\over 2}}$ becomes less than the
particle horizon scale, ${{2c} \over H_0} \Omega_0$ at $\Omega _0 < 0.85.$
Hence one expects a suppression of the low-order multipoles relative to the
higher order multipoles, over scales $l \simgt \Omega^{-1}.$  The detailed
shape of the predicted low multipole power spectrum is dominated by the
primordial entropy fluctuations and the details of how the fluctuation spectrum
is defined.

\item {d.}  Compton $y$-distortions of the CMB spectrum are inevitable in a BDM
universe at a level $y \sim 10^{-5}$ because of the early reionization.

\item {e.}  A peak in the matter power spectrum is inevitable at $100 - 300
\rm Mpc,$  corresponding to the maximum Jeans mass in the early universe.  This
could manifest itself as a source of a systematic, coherent large-scale flow
that is discrepant in direction with the CMB dipole, and possibly is aligned
with the CMB quadrupole (if such a quadrupole is indeed measured). Perhaps the
very large--scale flow $(\sim 800\rm\, km\, s^{-1})$ inferred from the dipole
moment of a sample of Abell clusters at a distance of $\sim 150\, h^{-1}\rm\,
Mpc$ \ref\laup$[\laup],$ if confirmed, would be best explained in such a model.

The major weakness in the BDM model arises from our poor understanding of star
formation in extreme environments, such as that of protogalaxies.  This applies
equally whether we wish to account for a level $\Omega_B \approx 0.02$ that
suffices to account for dark halos and to satisfy the primordial
nucleosynthesis constraint, or aim for the grander goal of $\Omega_B \approx
0.1$ in the cosmological setting.  The luminous regions of galaxies provide
$\Omega_{\ast} \approx 0.007$ in the form of known types of stars and gas.
There is not a unique prescription for arriving at this value of
$\Omega_{\ast}.$
 This is true regardless of whether the universal $\Omega$ is $0.1$ or $1.$
Thus it seems eminently plausible that BDM both does exist and should exist,
and dominate the known luminous matter content of the universe.

 Whether BDM
accounts for all of the matter in the universe is more problematical and
controversial.
Certainly, the trend towards a high $H_0$ pushes  one towards a low $\Omega$
universe, as does the most reliable and systematic-free, that is to say, the
most local, of the large-scale structure data.  Should BDM provide the
resolution to both the large-scale and the small-scale dark matter problems, it
is encouraging to note that we must be on the verge of detecting its elusive
signature.  BDM is (barely) alive and well.

 {\bf ACKNOWLEDGEMENTS}

I thank my students and colleagues for many discussions of topics covered in
these lectures. Wayne Hu provided the figures. This research has also been
supported in part by a grant from the N.S.F.


 \bigskip
  \parskip=0 truein
  \centerline {\bf REFERENCES}
  \medskip
\refs\waletal.  T. P. Walker, G. A. Steigman, D. N. Schramm, K. A. Olive and
H. S. Kang, {\sl Ap. J.} 376 (1991) 51.

\refs\pin. M. H. Pinsonneault, C. P. Deliyannis and P. Demarque, {\sl Ap. J.
Suppl.} 78 (1992) 181.

\refs\smil.  V. V. Smith, D. L. Lambert and P. E. Nissen, {\sl Ap. J.} 408
(1993) 262.

\refs\steetal. G. A. Steigman et al., {\sl Ap. J.} 415 (1993) L35.

\refs\walb.  T. P. Walker et al., {\sl Ap. J.} 413 (1993) 562.

 \refs\skr.  M. F. Skrutskie, M. A. Shure and S. Beckwith, {\sl Ap. J.} 299
(1985) 303.

 \refs\carf. C. Carignan and K. C. Freeman, {\sl Ap. J. Letters} 332 (1988)
L33.

\refs\dubc. J. Dubinski and R. Carlberg, {\sl Ap. J.} 378 (1991) 496.

 \refs\warq. M. S. Warren, P. J. Quinn, J. K. Salmon and W. H. Zurek, {\sl Ap.
J.} 399 (1992) 405.

\refs\moore. B. Moore, preprint (1994).

\refs\sacs. P. Sackett and L. S. Sparke, {\sl Ap. J.} 361 (1990) 408.

\refs\nelt. R. Nelson and S. Tremaine, private communication (1993).

\refs\begb. K. G. Begeman, {\sl Astr. Ap} 223 (1989) 47.

\refs\sana.  R. Sancisi and R. J. Allen,  {\sl Astr. Ap} 74 (1979) 73.

\refs\rup. M. P. Rupen, {\sl A. J.} 102 (1991) 48.

\refs\whi. S. D. M. White, {\sl Ap. J. Lett.} 294 (1985) L99.

\refs\binm. J. Binney and A. May, {\sl Mon. Not. Roy. astr. Soc.} 218 (1986)
743.

\refs\casv. S. Casertano and J. van Gorkom, {\sl A. J.} 101 (1991) 1231.

\refs\alcet. C. Alcock et al., {\sl Nature} 365 (1993) 621.

\refs\aubet. E. Aubourg et al., {\sl Nature} 365 (1993) 623.

\refs\pacet. A. Udalski et al., {\sl Acta Astr.} 418 (1993) 289.

\refs\mils. G. E. Miller and J. M. Scalo, {\sl Ap. J. Suppl.} 41 (1979) 513.

\refs\conv. P. S. Conti and Vacca, W. D., {\sl A. J.} 100 (1990) 431.

\refs\mez. R. G\"usten and P. G. Mezger, {\sl Vistas in Astr.} 26 (1983) 159.

\refs\scalo. J. Scalo, in {\sl Windows on Galaxies}, ed. G. Fabbiano et al.
(Kluwer) (1990) 125.

\refs\rie. G. H. Rieke et al., {\sl Ap. J.} 412 (1993) 99.

\refs\doam. J. S. Doane and W. G. Mathews, {\sl Ap. J.} 419 (1993) 573.

\refs\les. N. I. Gaffney, D. F. Lester and C. M. Telesco, {\sl Ap. J. Lett.}
407 (1993) L57.

\refs\arnr. M. Arnaud et al., {\sl Astr. Ap.} 254 (1992) 49.

\refs\renc. A. Renzini, L. Ciotti,  A. Dercole and S. Pellegrini, {\sl Ap. J.}
419 (1993) 52.

\refs\wolf. M. G. Wolfire and J. P. Cassinelli, {\sl Ap. J.} 319 (1987) 850.

\refs\lara. R. B. Larson, {\sl Mon. Not. Roy. astr. Soc} 214 (1985) 379.

\refs\worf. G. Worthey, S. M. Faber and J. J. Gonzalez, {\sl Ap. J.} 398 (1992)
69.

\refs\gus. B. Gustaffson et al., {\sl Astr. Ap.} 275 (1993) 101.

\refs\silkm. J. Silk, in {\it The Feedback of Chemical Evolution on the Stellar
Content of Galaxies}, ed. D. Alloin and G. Stasinska (Observatoire de Paris:
Meudon) (1992), 299.

\refs\barh. J. E. Barnes and L. Hernquist, {\sl Ann. Revs. Astr. Ap.} 30 (1992)
705.

\refs\lad. E. A. Lada, {\sl Ap. J. Lett} 393 (1992) L25.

\refs\myeetal. P. C. Myers et al., {\sl Ap. J.} 301 (1986) 398.

\refs\ladl. E. A. Lada and C. J. Lada, in  {\sl The Formation and Evolution of
Star Clusters}, ed. K. Janes, A. S. P. Conference Series, 13 (1991) 3.

\refs\silka. J. Silk, {\sl Austr. J. Phys.} 45 (1992) 437.

\refs\twa. B. A. Twarog, {\sl Ap. J.} 242 (1980) 242.

\refs\bar. D. C. Barry, {\sl Ap. J.} 334 (1988) 436.

\refs\nohs. H. R. Noh and J. Scalo, {\sl Ap. J.} 352 (1990) 605.

\refs\gal. J. S. Gallagher, D. A. Hunter and A. V. Tutukov, {\sl Ap. J.} 284
(1984) 544.

\refs\s. J. Silk,  in {\sl The Stellar  Populations of Galaxies}, IAU {\bf
149}, ed. B. Barbuy
and A. Renzini (Dordrecht: Kluwer) (1992) 367.

\refs\salp. E. E. Salpeter, {\sl Phys. Rep.} 227 (1993) 309.

\refs\Silk. J. Silk, {\sl Science}, 251 (1991) 537.

\refs\bes. M. S. Bessel, R. S. Sutherland and K. Ruan, {\sl Ap. J. Lett.} 383
(1991) L71.

\refs\lar.  R. B. Larson, {\sl Mon. Not. Roy. astr. Soc} 218 (1986) 409.

\refs\san. A. Sandage, {\sl Astr. Ap.} 161 (1986) 89.


  \refs\chit.  A. Chieffi and A. Tornambe, {\sl Ap. J.} 287 (1984)
745.

 \refs\fuji.  M. Y. Fujimoto, I. Iben, A. Chieffi and A. Tornambe, {\sl Ap. J.}
287 (1984) 749.

 \refs\ryu.  D. Ryu, K. A. Olive,  and J. Silk, {\sl Ap. J.} 353 (1990)
81.

\refs\moo. B. Moore, {\sl Ap. J. Lett.} 413 (1993) L93.

\refs\rixl. H.-W. Rix and G. Lake, {\sl Ap. J. Lett.}  417 (1993) L1.

\refs\burl. A.S. Burrows and J. Liebert, {\sl Rev. Mod. Phys} 65 (1993) 301.

\refs\bess. M. S. Bessel and G. Stringfellow, {\sl Ann. Revs. Astr. Ap.} 31
(1993) 433.

\refs\marb. G. W. Marcy, G. Basri and J. R. Graham, {\sl Ap. J. Lett.}, in
press (1993).

\refs\ric.  H.B Richer, and  G.G. Fahlman, {\sl Nature} 358 (1992) 383

\refs\huh. E. M. Hu, J.--S. Huang, G. Gilmore and L. L. Cowie, {\sl Nature},
submitted (1993).

\refs\ricc. H. B. Richer, D. R. Crabtree, A. C. Fabian and D. N. C. Lin, {\sl
A. J.} 105 (1993) 877.

\refs\white. D. A. White
 et al., {\sl Mon. Not. Roy. astr. Soc.} 252 (1991) 72.

\refs\richter. H. B. Richer et al.,  {\sl Ap. J.} 381 (1991) 147.

\refs\filet. A. Filippenko et al., {\sl A. J.} 104 (1992) 1543.

\refs\wans. B. Wang and J. Silk, {\sl Ap. J.} 406 (1993) 580.

\refs\tam. F. Tamanaha  et al., {\sl Ap. J.} 358 (1990) 164.

\refs\eic. D. Eichler and J. Silk, {\sl Science}, 257 (1992) 937.

\refs\bri. U. Briel, J. P. Henry and H. B\"ohringer, {\sl Astr. Ap.} 259 (1992)
L31.

\refs\wat. M. P. Watt  et al., {\sl Mon. Not. Roy. astr.Soc.} 258 (1992) 738.

\refs\eyl.  C. J. Eyles  et al., {\sl Ap. J.} 376 (1991) 23.

\refs\whif. D. A. White  et al., {\sl Mon. Not. Roy. astr.Soc.} 25 (1991) 72.

\refs\wolf. A. M. Wolfe in {\sl Relativistic Astrophysics and Cosmology}, ed.
C. W. Akerlof and M. A. Srednicki, {Ann. N. Y. Acad. Sci} 688 (1993) 281.

\refs\leqa. J. Lequeux, R. Allen and S. Guilloteau, {\sl Astr. Ap.} 280 (1993)
L23.

\refs\deg. E. J. de Geus et al., {\sl Ap. J. Lett} 413 (1993) L97.

\refs\derj. A. Derujula, Jetzer, P. and Masso, E., {\sl Astr. Ap.} 254 (1992)
99.

\refs\dols. A. Dolgov and J. Silk, {\sl Phys. Rev. D} 47 (1993) 4244.

\refs\yoks. J. Yokoyama and Y. Suto, {\sl Ap. J.} 379 (1991) 427.

\ref\husug

\refs\peeb. P. J. E. Peebles, {\sl Nature} 327 (1987) 210.

\refs\hus. W. Hu, D. Scott and J. Silk, {\sl Phys. Rev. D} in press (1994).

\refs\tegs. M. Tegmark and J. Silk, {\sl Ap. J.} in press (1994).

\refs\husug. W. Hu and N. Sugiyama, in preparation (1994).

\refs\laup. T. R. Lauer and M. Postman, preprint (1994).







\vfill\eject
\dimen10=\hsize \divide\dimen10 by 1
\vglue 3cm
\hglue 2cm

\centerline{$\vcenter{\box10}$}
\vglue 0cm
\vskip30pt
\setbox10\vbox{\overfullrule=0pt \parindent=0pt \hsize=\dimen10
Figure 1.
The matter transfer function for baryonic dark matter--dominated cosmological
models $[\husug].$ For each combination of $\Omega$ and $h,$ results are shown
for several epochs of reionization.}
\centerline{$\vcenter{\box10}$}

\vfill\eject
\dimen10=\hsize \divide\dimen10 by 1
\hglue 2cm
\vglue 3cm

\centerline{$\vcenter{\box10}$}
\vglue 0cm
\vskip30pt
\setbox10\vbox{\overfullrule=0pt \parindent=0pt \hsize=\dimen10
Figure 2.
The velocity correlation function for baryonic dark matter--dominated
cosmological models $[\husug].$ For $\Omega=0.2$ and $h=1,$ results are shown
with normalization $\sigma_8=1$ for  reionization at $z=1000$ compared to no
reionization, and for $n=-1$ and $n=0.$. For comparison, also shown is the
velocity correlation function for unbiased CDM.}
\centerline{$\vcenter{\box10}$}

\end